\documentclass{article}
\usepackage{amsfonts}
\usepackage{amssymb}
\usepackage{graphicx}
\usepackage{amsmath}

\input{tcilatex}
\begin{document}

\title{Model of a two--transverse mode laser with injected signal}
\author{Germ\'{a}n J. de Valc\'{a}rcel, Eugenio Rold\'{a}n, \\
Departament d'\`{O}ptica, Universitat de Val\`{e}ncia, \\
Dr. Moliner 50, 46100--Spain \and and, Kestutis Staliunas \\
Departament de F\'{\i}sica i Enginyeria Nuclear, \\
Universitat Polit\`{e}cnica de Catalunya, \\
Colomm 11, 08222--Terrassa, Spain}
\maketitle

\begin{abstract}
We derive a simple model for a two transverse mode laser (that considers the
TEM$_{00}$ and TEM$_{10}$ modes) in which an injected signal with the shape
of the TEM$_{10}$ mode but a frequency close to that of the TEM$_{00}$ mode
is injected.
\end{abstract}

\date{}

Here we derive a model for a two--transverse mode laser with injected
signal. More specifically, we do consider an incoherently pumped
homogeneously broadened laser, in a ring cavity configuration in which a
particular signal is injected from the outside. We assume that the laser
works reasonably close to resonance with the TEM$_{00}$ mode, but the
injected field has not the shape of the TEM$_{00}$ mode but the shape of the
TEM$_{10}$ mode. The frequency of the injected field is however assumed to
be close to that of the TEM$_{00}$ mode. We first develop a two--mode model
which is valid for any value of the decay rates. Later on, we particularize
this model to class--A lasers, i.e., lasers in which the material variables
decay much faster than the intracavity field. We end up with a single
complex equation for the TEM$_{00}$ mode which is of the Stuart--Landau type.

\section{General model}

The starting point are the laser equations for a large Fresnel number
ring--cavity with spherical mirrors filled with a collection of
homogeneously broadened two--level atoms that are incoherently pumped. These
equations can be found, e.g., in \cite{Lugiato90,Brambilla94}. To this model
equations we add an injected signal with arbitrary spatial structure. In the
single longitudinal mode and uniform field approximations these equations
can be written in dimensionless form as 
\begin{subequations}
\label{complet}
\begin{align}
\frac{\partial }{\partial t}F\left( \mathbf{r},t\right) & =-\left( 1+i\theta
\right) F+ia\mathcal{L}F+2CP+F_{\mathrm{in}}\left( \mathbf{r}\right) , \\
\frac{\gamma _{\mathrm{c}}}{\gamma _{\perp }}\frac{\partial }{\partial t}%
P\left( \mathbf{r},t\right) & =-\left( 1+i\Delta \right) P+FD, \\
\frac{\gamma _{\mathrm{c}}}{\gamma _{||}}\frac{\partial }{\partial t}D\left( 
\mathbf{r},t\right) & =-D+\chi -\func{Re}\left( F^{\ast }P\right) .
\end{align}%
In these equations $F$ and $P$ are proportional to the slowly varying
complex amplitudes of the laser electric field and the medium polarization,
and $D$ is proportional to the atomic inversion, which decay at rates $%
\gamma _{\mathrm{c}}$, $\gamma _{\perp }$, and $\gamma _{||}$, respectively.
Time $t$ is measured in units of $\gamma _{\mathrm{c}}^{-1}$. The external
monochromatic injection is defined by its complex amplitude, proportional to 
$F_{\mathrm{in}}\left( \mathbf{r}\right) $, and by its frequency $\omega _{%
\mathrm{in}}$, which serves as the frequency frame in which the equations
have been written, so the detunings read 
\end{subequations}
\begin{equation}
\theta =\frac{\omega _{\mathrm{c}}-\omega _{\mathrm{in}}}{\gamma _{\mathrm{c}%
}},\;\Delta =\frac{\omega _{\mathrm{a}}-\omega _{\mathrm{in}}}{\gamma
_{\perp }},
\end{equation}%
where $\omega _{\mathrm{c}}$ is the cavity frequency of the fundamental, TEM$%
_{00}$, transverse mode, and $\omega _{\mathrm{a}}$ is the atomic frequency. 
$C$ is the usual cooperativity parameter and $\chi $ sets the value of the
inversion in the absence of fields.

Finally operator $\mathcal{L}=\left( \frac{1}{4}\nabla ^{2}-r^{2}+1\right) $
accounts for the modal structure of the resonator, with $\nabla ^{2}=\left(
\partial ^{2}/\partial x^{2}+\partial ^{2}/\partial y^{2}\right) $ the
Laplacian acting on the transverse coordinates $\mathbf{r}=\left( x,y\right) 
$, which are measured in units of the TEM$_{00}$ beam waist radius, and $a$
is the transverse mode spacing (measured in units of $\gamma _{\mathrm{c}}$%
). We note that the set of Hermite-Gauss (TEM$_{mn}$) modes%
\begin{equation}
\Psi _{m,n}\left( \mathbf{r}\right) =\frac{1}{\sqrt{2^{m+n-1}m!n!\pi }}%
H_{m}\left( \sqrt{2}x\right) H_{n}\left( \sqrt{2}y\right) e^{-r^{2}},
\end{equation}%
are eigenfunctions of $\mathcal{L}$,%
\begin{equation}
\mathcal{L}\Psi _{m,n}\left( \mathbf{r}\right) =-\left( m+n\right) \Psi
_{m,n}\left( \mathbf{r}\right) .
\end{equation}%
The Hermite-Gauss modes verify the orthonormality condition%
\begin{equation}
\dint \mathrm{d}^{2}\mathbf{r}\ \Psi _{m,n}\left( \mathbf{r}\right) \Psi
_{m^{\prime },n^{\prime }}\left( \mathbf{r}\right) =\delta _{m,m^{\prime
}}\delta _{n,n^{\prime },}
\end{equation}%
and form a basis for functions defined on the transverse plane, hence any
function $G\left( \mathbf{r}\right) $ can be expanded into it as%
\begin{equation}
G\left( \mathbf{r},t\right) =\dsum\limits_{m,n}g_{m,n}\left( t\right) \Psi
_{m,n}\left( \mathbf{r}\right) ,
\end{equation}%
where $g_{m,n}\left( t\right) $ are the modal coefficients.

The above is the starting model from which we pass to derive a two--mode
model.

\section{Two mode class C laser model}

We consider the situation in which the (normalized) frequency spacing
between the transverse modes, $a$, which only depends on the cavity
geometry, is large ($a\gg 1$) while the (normalized) detunings $\theta $,
and $\Delta $ are, at most, of order $1$. Without injection, it is evident
that under the above conditions only the fundamental, Gaussian TEM$_{00}$
mode will be excited above threshold. Nevertheless the injected field is
chosen to have the shape of the TEM$_{10}$ mode,%
\begin{equation}
F_{\mathrm{in}}\left( \mathbf{r}\right) =f_{\mathrm{in}}\Psi _{1,0}\left( 
\mathbf{r}\right) ,
\end{equation}%
with $f_{\mathrm{in}}$ the injected field amplitude, which we take as a
constant parameter. Hence, apart from the TEM$_{00}$ mode, the TEM$_{10}$
mode will be oscillating as well, and then it is feasible to approximate the
total field as%
\begin{equation}
F\left( \mathbf{r},t\right) =f_{0}\left( t\right) \Psi _{0,0}\left( \mathbf{r%
}\right) +f_{1}\left( t\right) \Psi _{1,0}\left( \mathbf{r}\right) ,
\end{equation}%
at least when the system is not very far from threshold. Of course this
approximation will be invalid if the injected field has a frequency which is
close to that of the TEM$_{10}$ mode, and in this case more mode families
should be taken into account. We shall remind later on the smallness of the
injected signal detuning as compared with the frequency spacing between the
transverse modes. In a similar way we approximate 
\begin{subequations}
\begin{align}
P\left( \mathbf{r},t\right) & =p_{0}\left( t\right) \Psi _{0,0}\left( 
\mathbf{r}\right) +p_{1}\left( t\right) \Psi _{1,0}\left( \mathbf{r}\right) ,
\\
D\left( \mathbf{r},t\right) & =d_{0}\left( t\right) \Psi _{0,0}\left( 
\mathbf{r}\right) +d_{1}\left( t\right) \Psi _{1,0}\left( \mathbf{r}\right) .
\end{align}

By substituting the above expansions into the original model equations, and
projecting onto the different modes, the following evolution equations for
the amplitudes are found 
\end{subequations}
\begin{subequations}
\label{completC}
\begin{align}
\frac{\mathrm{d}}{\mathrm{d}t}f_{0}\left( t\right) & =-f_{0}-i\theta
f_{0}+2Cp_{0}, \\
\frac{\mathrm{d}}{\mathrm{d}t}f_{1}\left( t\right) & =-f_{1}-i\left( \theta
+a\right) f_{1}+2Cp_{1}+f_{\mathrm{in}}, \\
\frac{\gamma _{\mathrm{c}}}{\gamma _{\perp }}\frac{\mathrm{d}}{\mathrm{d}t}%
p_{0}\left( t\right) & =-p_{0}-i\Delta p_{0}+\eta f_{0}d_{0}+\frac{2}{3}\eta
f_{1}d_{1}, \\
\frac{\gamma _{\mathrm{c}}}{\gamma _{\perp }}\frac{\mathrm{d}}{\mathrm{d}t}%
p_{1}\left( t\right) & =-p_{1}-i\Delta p_{1}+\frac{2}{3}\eta \left(
f_{0}d_{1}+f_{1}d_{0}\right) , \\
\frac{\gamma _{\mathrm{c}}}{\gamma _{||}}\frac{\mathrm{d}}{\mathrm{d}t}%
d_{0}\left( t\right) & =-d_{0}+\chi _{0}-\eta \func{Re}\left( f_{0}^{\ast
}p_{0}\right) -\frac{2}{3}\eta \func{Re}\left( f_{1}^{\ast }p_{1}\right) , \\
\frac{\gamma _{\mathrm{c}}}{\gamma _{||}}\frac{\mathrm{d}}{\mathrm{d}t}%
d_{1}\left( t\right) & =-d_{1}-\frac{2}{3}\eta \func{Re}\left( f_{0}^{\ast
}p_{1}+f_{1}^{\ast }p_{0}\right) .
\end{align}%
where $\eta =\frac{2\sqrt{2}}{3\sqrt{\pi }}$.

These are our two-transverse mode class--C model equations. It is convenient
to recall that the range of validity of these equations is limited to large
values of the transverse mode spacing ($a\gg 1$) and to small values of the
detunings $\theta $ and $\Delta $ (at most of order $a^{0}$), and not large
pump nor large injected signal.

\section{Two-mode Class A laser model}

In order to simplify the problem as much as possible we consider a class A
laser, defined by 
\end{subequations}
\begin{equation}
\frac{\gamma _{\mathrm{c}}}{\gamma _{\perp }},\frac{\gamma _{\mathrm{c}}}{%
\gamma _{||}}\ll 1,
\end{equation}%
and consequently we eliminate adiabatically the medium variables.
Furthermore the condition $a\gg 1$ allows eliminating the modal amplitude $%
f_{1}$ as well. Finally, we will consider that the laser is operated close
to threshold, as commented, and use a cubic approximation in the fields as
usual. With all these elements one can express the model variables $p_{0}$, $%
p_{1}$, $d_{0}$, $d_{1}$, and $f_{1}$ in terms of the fundamental mode
amplitude $f_{0}$ and of the model parameters.

Substitution of these expressions into the equation for $f_{0}$ yields
finally%
\begin{equation}
\frac{\mathrm{d}}{\mathrm{d}t}A\left( t\right) =\alpha A+\beta A^{\ast
}-\left( 1-i\Delta \right) \left\vert A\right\vert ^{2}A,  \label{ecfin}
\end{equation}%
where the relation between the field $A$ and the Gaussian mode amplitude $%
f_{0}$ is%
\begin{equation}
A\left( t\right) =\eta f_{0}\left( t\right) ,
\end{equation}%
and the complex parameters $\alpha $ and $\beta $ in Eq. (\ref{ecfin}) are
given by%
\begin{eqnarray}
\func{Re}\left( \alpha \right)  &=&\frac{r}{1+\Delta ^{2}}-1-\frac{\left(
11-\Delta ^{2}\right) a^{2}E^{2}}{5\left( 1+\Delta ^{2}\right) \left(
a+\theta \right) ^{2}}, \\
\func{Im}\left( \alpha \right)  &=&-\frac{r\Delta }{1+\Delta ^{2}}-\theta +%
\frac{12\Delta a^{2}E^{2}}{5\left( 1+\Delta ^{2}\right) \left( a+\theta
\right) ^{2}}, \\
\beta  &=&\frac{a^{2}E^{2}}{5\left( 1+\Delta ^{2}\right) \left( a+\theta
\right) ^{2}}\left( 5+\Delta ^{2}-4i\Delta \right) ,
\end{eqnarray}%
where 
\begin{subequations}
\begin{eqnarray}
E &=&\frac{2}{3}\sqrt{\frac{6}{5}}\frac{\eta }{a}f_{\mathrm{in}}, \\
r &=&2C\eta \chi _{0}.
\end{eqnarray}

Equation (\ref{ecfin}) constitutes the final model. We note that $A\left(
t\right) $ is proportional to the amplitude $f_{0}\left( t\right) $ of the
TEM$_{10}$ mode, and that $E$ is a free parameter proportional to the
amplitude of the external injection, $f_{\mathrm{in}}$. The remaining
parameters are the detunings $\theta $ and $\Delta $, and the pump parameter 
$r$, dependent on the usual pump parameter $2C$: It is easy to derive that
the threshold value of the free--running laser (with injection $E=0$) is
given by $r=1+\Delta ^{2}$, i.e. by 
\end{subequations}
\begin{equation}
2C=\frac{1+\Delta ^{2}}{\eta \chi _{0}}.
\end{equation}

Let us remind that in the derivation we have assumed that the fields are
weak, thus implying that $r$ is close its threshold value, that the decay
rates of the material variables are much larger than that of the intracavity
field, and that the intermode spacing is large compared to the cavity decay
rate ($a>>1$). We have also assumed that $2\theta <a$ as for larger cavity
detunings it does not make sense the assumption that the laser is working,
basically, in the TEM$_{00}$ mode.

We find it worth writing the above Stuart--Landau equation in the special
case $\Delta =0$ (i.e., $\omega _{\mathrm{in}}=\omega _{\mathrm{a}}$), as
this can be easily done in the laboratory. In this case the equation
simplifies to%
\begin{equation}
\frac{\mathrm{d}}{\mathrm{d}t}A\left( t\right) =\left( r-1-\frac{11}{5}\frac{%
a^{2}E^{2}}{\left( a+\theta \right) ^{2}}-i\theta \right) A+\frac{a^{2}E^{2}%
}{\left( a+\theta \right) ^{2}}A^{\ast }-\left\vert A\right\vert ^{2}A.
\end{equation}%
Notice that the infinite resonances appearing in parameters $\alpha $ and $%
\beta $ for $\theta =-a$ is an artifact, as in the derivation we have
implicitely assumed that the cavity detuning is smaller than half the
intermode spacing, i.e., $2\theta <a$.

\end{document}